\documentclass[10pt, conference, table]{IEEEtran}
\IEEEoverridecommandlockouts

\usepackage[caption=false,font=footnotesize]{subfig}
\usepackage{cite}
\usepackage{dsfont}
\usepackage{amsmath,amssymb,amsfonts}
\usepackage{algorithmic}
\usepackage{graphicx}
\usepackage{textcomp}
\usepackage{hyperref}
\usepackage{enumitem}
\usepackage{booktabs}
\usepackage{graphicx}
\usepackage{arydshln} 
\usepackage{multirow}
\usepackage{pifont}
\usepackage{cleveref}
\usepackage{array}
\usepackage{xcolor}
\usepackage{fancyhdr}

\usepackage{xcolor}
\def\BibTeX{{\rm B\kern-.05em{\sc i\kern-.025em b}\kern-.08em
    T\kern-.1667em\lower.7ex\hbox{E}\kern-.125emX}}

\usepackage{fancyhdr}
\fancypagestyle{plain}{%
  \fancyhf{}%
  \fancyfoot[C]{\footnotesize Accepted to Swedish National Computer Networking Workshop (SNCNW) 2026}%
}

\begin{document}

\title{Rethinking IoT Intrusion Detection: Augmenting Routing Metrics with Radio Features}

\author{\IEEEauthorblockN{Yichang Sun, Andreas Johnsson, and Sourasekhar Banerjee, }
\IEEEauthorblockA{Uppsala University, Department of Information Technology, Sweden\\
Email: \emph{\{yichang.sun.3197, andreas.johnsson, sourasekhar.banerjee\}@it.uu.se}  
}

}

\maketitle
\thispagestyle{fancy}
\begin{abstract}
Machine learning-based intrusion detection systems (IDS) for RPL-based IoT networks often rely solely on routing-layer features, which provide only a partial view of network behaviour. In this work, we investigate whether incorporating Transmit (TX) and Receive (RX) radio features alongside the standard RPL feature set can improve detection performance in an LSTM-based IDS. We evaluate the proposed approach across three different attack types, namely DIS-Flooding, Local Repair, and  Worst Parent under varying network sizes. The results show that incorporating TX and RX improves the IDS's overall detection performance by up to $\sim4\%$ in F1-score compared with using routing-layer features alone, with the most notable gain observed for the Worst Parent attack.
\end{abstract}


\section{Introduction}
The Internet of Things (IoT) has become a key enabler of modern digital infrastructure across smart homes, healthcare, and industrial systems. Many such deployments operate over low-power and lossy networks, where the IPv6 Routing Protocol for Low-Power and Lossy Networks (RPL) provides scalable and efficient routing. However, RPL remains vulnerable to a broad range of routing attacks that can disrupt topology formation, degrade packet delivery, increase control overhead, and exhaust node resources. This problem motivates the need for robust intrusion detection systems (IDS) in IoT networks \cite{raoof2018routing,prajapati2025comprehensive}.
Machine learning-based IDS has shown promising results by learning traffic patterns associated with normal and malicious behaviour \cite{azumah2021deep}, with LSTM models being particularly effective due to their ability to capture temporal dependencies in sequential network data. Prior works \cite{kaveh2024impact, kaveh2025factors, banerjee2026quantifying} have demonstrated that RPL control traffic statistics collected at the sink node can be used to train effective IDS models, though detection performance remains sensitive to attack type, topology, and network size. However, existing approaches largely rely on routing-layer features alone \cite{azumah2021deep, banerjee2026quantifying}. Previous studies \cite{przybocki2023analysis} have also shown that RPL-oriented denial-of-service attacks can substantially affect device-level communication and resource-consumption behaviour, including transmit (TX) and receive (RX) signal activity, CPU time, and power draw \cite{smith2020battery,przybocki2023analysis}. Many RPL-based attacks affect not only routing behaviour, but also radio activity, meaning this routing-centric view may provide an incomplete representation of attack impact. Moreover, cross-layer intrusion detection approaches have begun to show that incorporating lower-layer features alongside routing statistics can improve detection performance in RPL-based IoT networks \cite{canbalaban2020cross}.
Motivated by the observation that existing IDS approaches \cite{banerjee2026quantifying, kaveh2025factors} rely primarily on routing-layer features, this paper investigates whether extending the RPL feature set with Transmit (TX) and Receive (RX) radio features can improve intrusion detection performance. TX and RX capture node-level communication behaviour that routing-layer statistics alone cannot observe. For example, DIS-Flooding forces neighbouring nodes to reset their trickle timers and respond with DIO broadcasts. That increases the RX count at those nodes beyond normal operational levels. Since these attack-induced changes in radio activity evolve over time, we employ an LSTM-based classifier to capture temporal dependencies in sequential network data.
\noindent\textit{\textbf{Contributions:}} We extend the baseline RPL feature set \cite{banerjee2026quantifying, kaveh2025factors} with TX and RX radio features and evaluate their impact on LSTM-based intrusion detection in RPL-based IoT networks. We further compare multiple feature configurations across three attack types, three behavioural variants (base, on-off, and gradual change), and different network sizes (5, 10, 15, 20) to assess the contribution of radio features relative to the RPL-only baseline within the current experimental pipeline.

\begin{figure}[!t]
    \centering
    \includegraphics[width=0.9\linewidth]{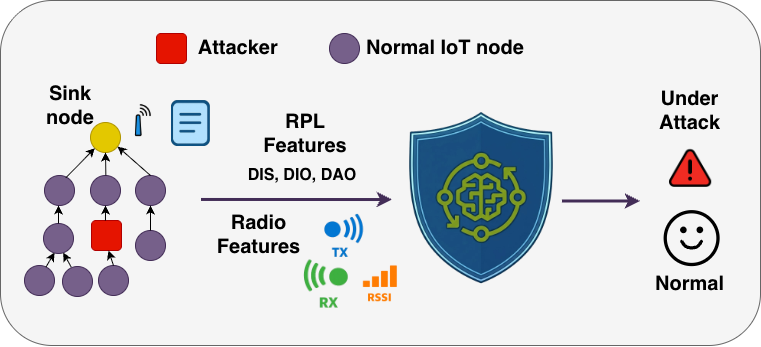}
    \caption{Proposed LSTM-based IDS framework for RPL-based IoT networks}
    \label{fig: framework}
\end{figure}

\section{System Model and Problem Formulation}
In \Cref{fig: framework}, we illustrate the proposed LSTM-based IDS framework for an RPL-based IoT network. It consists of $N$ nodes: a set of normal IoT nodes forming a DODAG, a single sink node at the root responsible for data collection, and one attacker injecting malicious behaviour into the network. The sink node passively aggregates two categories of features from the network. The first category comprises RPL features, including Rank, DIS, DIO, and DAO control message statistics that capture routing-layer behaviour. The second category consists of radio features, comprising TX and RX activity that reflect node-level communication patterns. The IDS processes these combined features to classify network behaviour as either \textit{Normal} or \textit{Under Attack}.

The problem can therefore be formulated as follows. At each time step $t$, the sink node observes a feature vector composed of RPL control traffic statistics:
\[
\mathbf{x}_t^{\text{RPL}} = [\text{Rank},\, \text{DIS},\, \text{DIO},\, 
\text{DAO}, \cdots]
\]
We extend this baseline representation by incorporating radio features:
\[
\mathbf{x}_t^{\text{aug}} = [\mathbf{x}_t^{\text{RPL}} \;\|\; \text{TX}_t,\, 
\text{RX}_t]
\]
where $\|$ denotes concatenation. Given a sequence $\mathbf{X} = [\mathbf{x}_1, \ldots, \mathbf{x}_T]$, an LSTM-based classifier $f_\theta$ is trained to map  the input to a label $y \in \{0, 1 \}$, corresponding to normal behaviour or not. The objective is to find an augmented feature representation that maximizes the detection performance of the LSTM-based IDS:
\[
\mathbf{x}_t^{*} = \arg\max_{\mathbf{x}_t \in \mathbf{x}^{\text{aug}}_t} \; 
\mathcal{P}(f_\theta(\mathbf{X}), \mathbf{y})
\]
where $\mathbf{y}$ is the true label, 
and $\mathcal{P}$ denotes detection performance measured by F1-score across 
attack types, their behavioural variants, and network sizes.

\section{Experimental Setup} \label{sec: experimental_setup}

\noindent\textbf{Datasets}: We extend the publicly available simulation scripts from \cite{uu-core_IoT_Attacks_IDS}, which originally collect only RPL control traffic statistics, by incorporating additional scripts to extract TX and RX radio features from the Cooja/Contiki-NG \cite{finne2021multi} environment. In the current implementation, these radio features are derived from log outputs that capture node-level radio activity during the simulation. Using this extended pipeline, we simulate three RPL-based attack types, namely DIS-Flooding, Local Repair, and  Worst Parent, each evaluated under three behavioural variants, i.e., base, on--off, and gradual change, across network sizes of 5, 10, 15, and 20 nodes. For each configuration, 20 independent simulation runs are performed, each producing one CSV file containing the recorded RPL and radio statistics. These logs are subsequently processed into window-level features for IDS training and evaluation.

\noindent\textbf{Features:} The RPL feature vector $\mathbf{x}_t^{\text{RPL}} \in \mathbb{R}^{14}$ is 
constructed from the mean ($\mu$) and standard deviation ($\sigma$) of seven routing-layer statistics observed at the sink node:
\begin{multline*}
\mathbf{x}_t^{\text{RPL}} = \bigl[\,
\mu_{\text{Rank}},\, \sigma_{\text{Rank}},\,
\mu_{\text{DIS}}^{\text{s}},\, \sigma_{\text{DIS}}^{\text{s}},\,
\mu_{\text{DIS}}^{\text{r}},\, \sigma_{\text{DIS}}^{\text{r}},\, \\
\mu_{\text{DIO}}^{\text{s}},\, \sigma_{\text{DIO}}^{\text{s}},\,
\mu_{\text{DIO}}^{\text{r}},\, \sigma_{\text{DIO}}^{\text{r}},\,
\mu_{\text{DAO}}^{\text{r}},\, \sigma_{\text{DAO}}^{\text{r}},\,
\mu_{\text{tots}},\, \sigma_{\text{tots}}
\,\bigr]
\end{multline*}
where superscripts $\text{s}$ and $\text{r}$ denote message sent and received counts respectively, and $\text{tots}$ denotes the total number of messages observed at the sink node. To extend this representation, we incorporate radio features intended to reflect node-level transmission and reception activity:
\[
\mathbf{x}_t^{\text{radio}} = \bigl[\,
\mu_{\text{TX}},\, \sigma_{\text{TX}},\,
\mu_{\text{RX}},\, \sigma_{\text{RX}}
\,\bigr] \in \mathbb{R}^{4}
\]

The full augmented feature vector is formed by concatenating the RPL and 
radio feature sets:

\[
\mathbf{x}_t^{\text{aug}} = [\,\mathbf{x}_t^{\text{RPL}} \;\|\; 
\mathbf{x}_t^{\text{radio}}\,] \in \mathbb{R}^{18}
\]

\noindent\textbf{Model:} We employ an LSTM-based model for binary classification of network behaviour into normal ($0$) and attack ($1$) classes. Given an input sequence $\mathbf{X} = [\mathbf{x}_1, \mathbf{x}_2, \ldots, \mathbf{x}_T] \in \mathbb{R}^{T \times d}$, where $T$ is the window length and $d$ is the input dimension, the LSTM computes a hidden state at each time step:
\[\mathbf{h}_t = \text{LSTM}(\mathbf{x}_t, \mathbf{h}_{t-1})\]

The final hidden state $\mathbf{h}_T \in \mathbb{R}^{H}$ is passed through a fully connected output layer with a sigmoid activation to produce the attack probability:

\[
\hat{p} = \sigma(\mathbf{W}\mathbf{h}_T + \mathbf{b}), \quad \hat{p} \in (0, 1)
\]

where $\mathbf{W}$ and $\mathbf{b}$ are learnable parameters and 
$\sigma(\cdot)$ is the sigmoid function. The predicted class is obtained as $y^* = \mathds{1}[\hat{p} > 0.5]$, corresponding to either normal or attack traffic. The input dimension $d$ depends on the feature configuration: $d = 14$ for the RPL-only baseline and $d = 18$ for the augmented setting. The model is configured with a hidden size of $H = 10$, a dropout rate of $0.3$, and is trained using the binary cross-entropy loss:
\begin{equation*}
\mathcal{L} = -\frac{1}{N}\sum_{i=1}^{N} \bigl[ y_i \log(\hat{p}_i) + 
(1 - y_i) \log(1 - \hat{p}_i) \bigr]
\end{equation*}

where $y_i \in \{0, 1\}$ is the true label and $\hat{p}_i$ is the predicted 
probability of the attack class for sample $i$.

\noindent\textbf{IDS learning and evaluation:} 
An independent LSTM model is trained and evaluated for each experimental configuration, defined by a combination of attack type, behavioural variant, and network size. This results in a total of $3 \times 3 \times 4 = 36$ configurations, 
covering three attack types (DIS-Flooding, Local Repair, and  Worst Parent), three behavioural variants (base, on--off, and gradual change), and four network sizes ($N \in \{5, 10, 15, 20\}$).

The sequential input samples are generated using a sliding-window approach with a window size of 10 and a step size of 3. The model is trained using the AdamW optimizer with a learning rate of $0.001$ and the binary cross-entropy loss. The batch size is set to 128, and training runs for 50 epochs. The best model checkpoint is selected based on the lowest validation loss. The IDS is evaluated by the F1-Score.

\section{Results}
\begin{figure}[!t]
    \centering
    \subfloat[DIS-Flooding]{
        \includegraphics[width=0.45\textwidth]{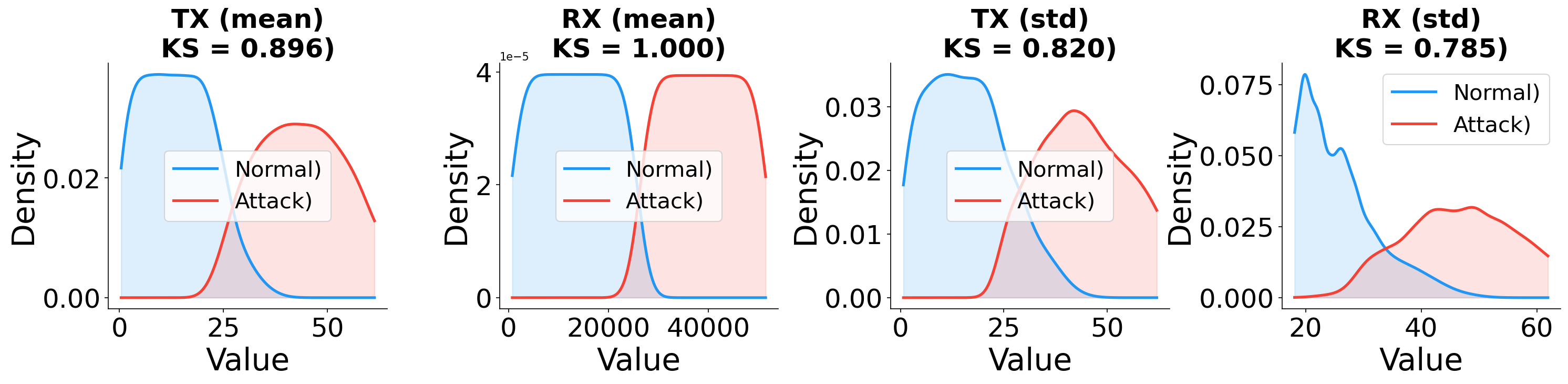}
        \label{fig:second}
    }\hfil
    \subfloat[Local Repair]{
        \includegraphics[width=0.45\textwidth]{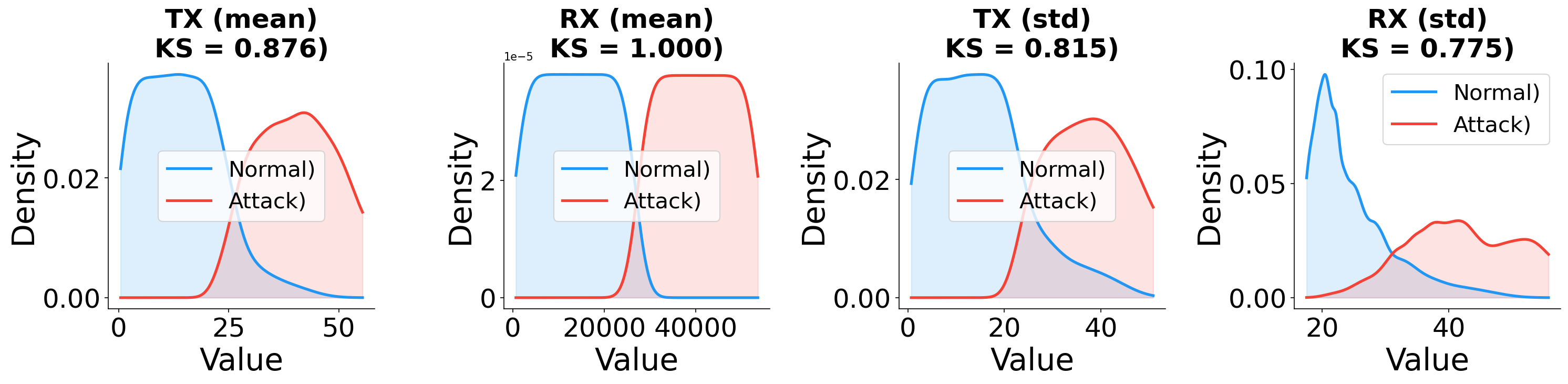}
        \label{fig:third}
    }\hfil
    \subfloat[Worst Parent]{
        \includegraphics[width=0.45\textwidth]{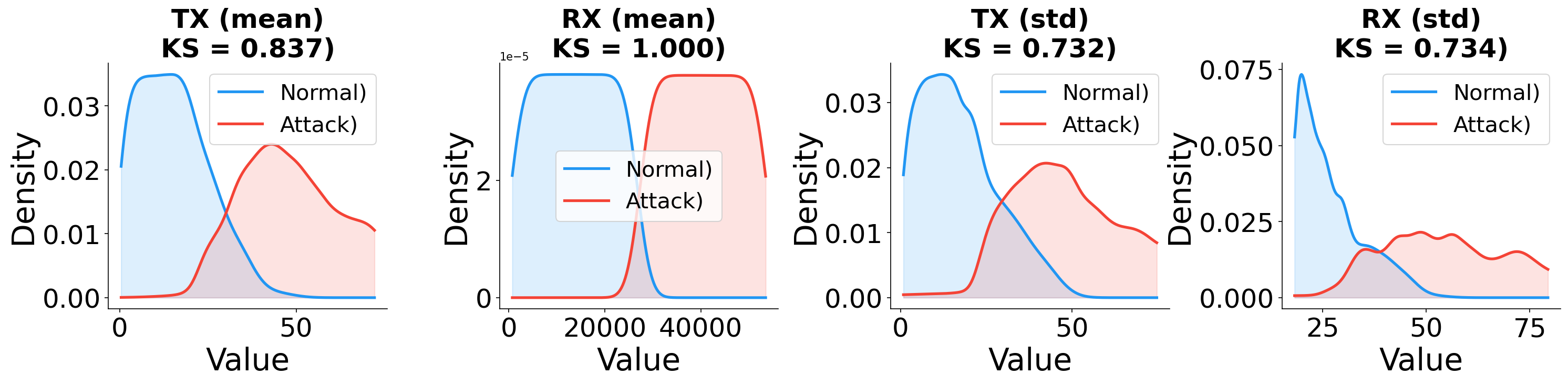}
        \label{fig:fourth}
    }
    \caption{Feature distributions of TX and RX (mean and std) under normal and attack conditions for three RPL-based attacks in a 20-node network (base variant), averaged over 20 independent simulation runs. The Kolmogorov--Smirnov (KS) statistic quantifies the separation between normal and attack distributions, where a higher value indicates greater statistical divergence between the two classes.}
    \label{fig:dist}
\end{figure}

\subsection{Radio Feature Analysis}

\Cref{fig:dist} shows the kernel density distributions of the mean and standard deviation of TX and RX under normal and attack conditions for the base variant of three RPL-based IoT attacks (DIS-Flooding, Local Repair, and Worst Parent) in a 20-node network. The Kolmogorov--Smirnov (KS) statistic \cite{massey1951kolmogorov} quantifies the separation between normal and attack distributions, with a higher value indicating greater separability.

Across all three attack types, RX mean consistently achieves a perfect KS score of 1.000, indicating complete distributional separation between normal and attack traffic. TX mean also exhibits strong separation, with KS values ranging from 0.837 (Worst Parent) to 0.896 (DIS-Flooding). The standard deviation features, TX std and RX std, show comparatively lower but still substantial KS values, ranging from 0.732 to 0.820. The KS statistic ranges from 0 to 1, where a value closer to 1.0 indicates greater statistical divergence between the normal and attack class distributions. Since all evaluated TX and RX features achieve KS scores above 0.7, the normal and attack distributions are well-separated, reflecting how attack conditions shift wireless transmission and reception activity at the node level beyond normal operational levels. This strong distributional separability suggests that incorporating TX and RX into IDS training will provide the LSTM  classifier with discriminative signals that are not captured by RPL control-message statistics alone. 
\begin{figure*}[!t]
    \centering
    \subfloat[DIS-Flooding]{
        \includegraphics[width=0.31\textwidth]{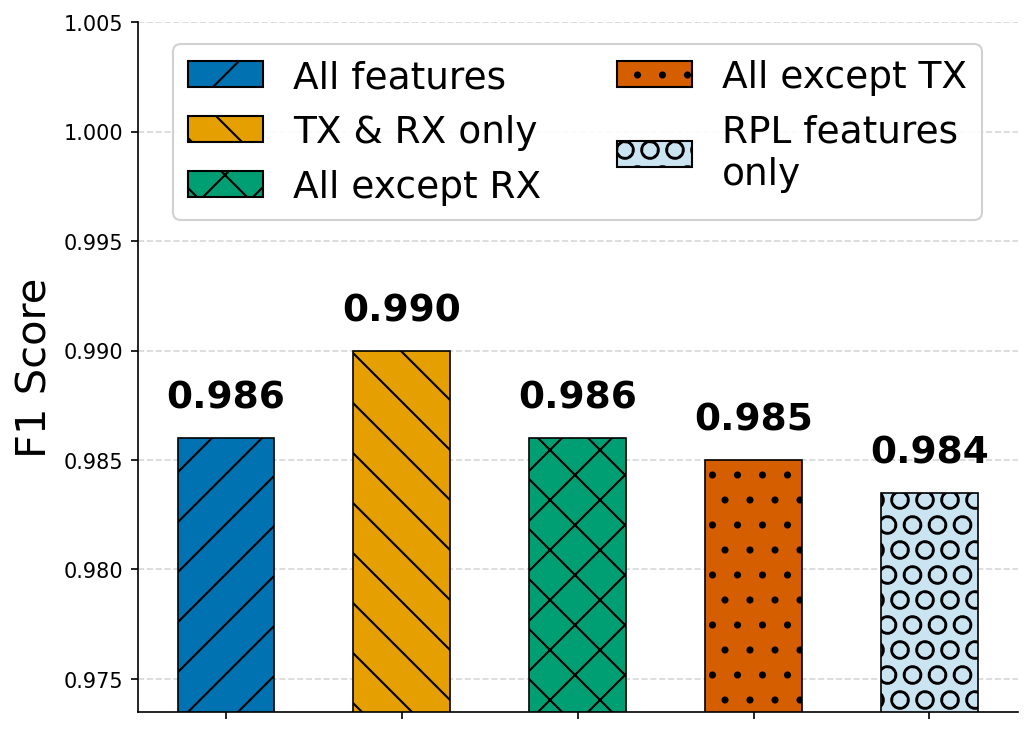}
        \label{fig:f1_second}
    }\hfil
    \subfloat[Local Repair]{
        \includegraphics[width=0.31\textwidth]{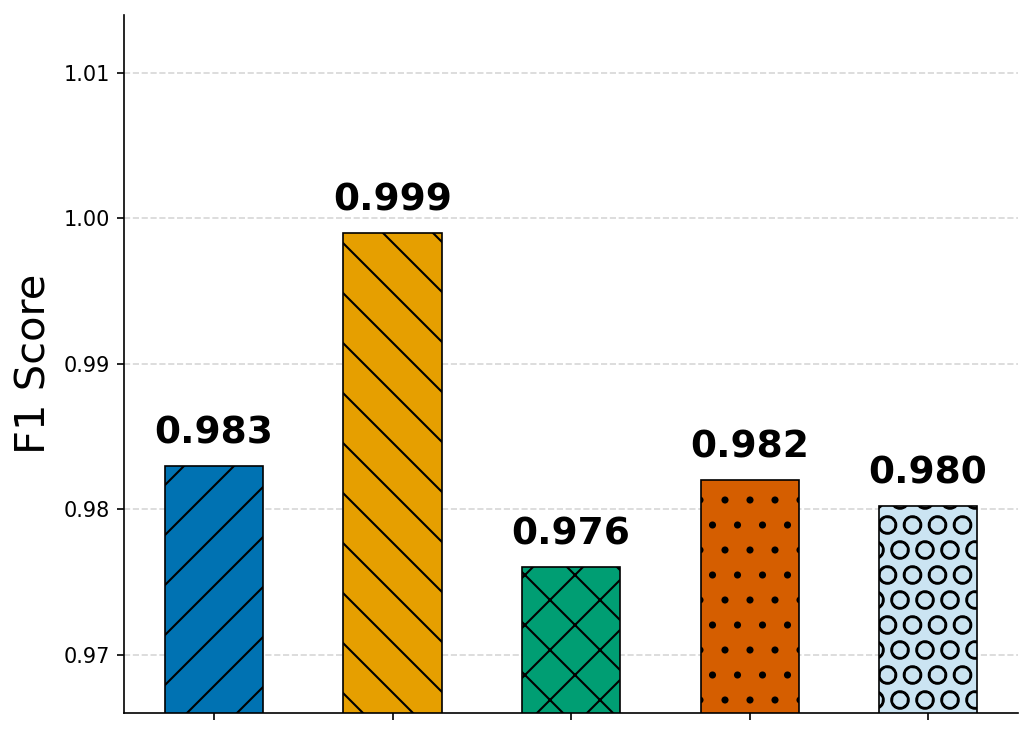}
        \label{fig:f1_third}
    }\hfil
    \subfloat[Worst Parent]{
        \includegraphics[width=0.31\textwidth]{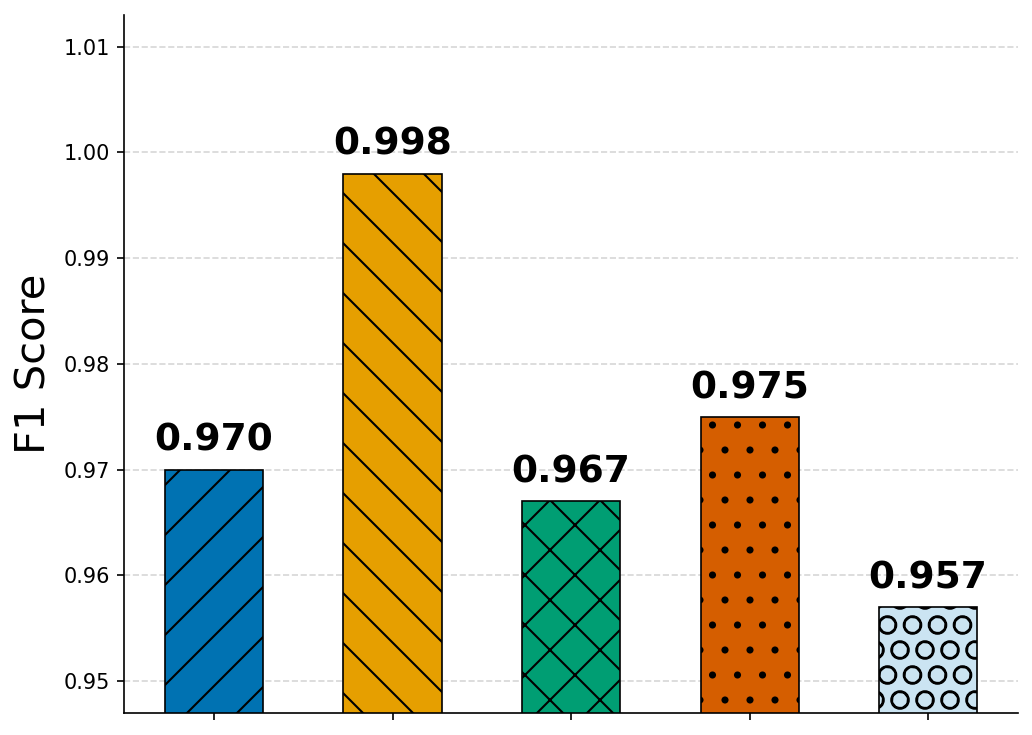}
        \label{fig:f1_fourth}
    }
    \caption{Mean F1 scores of different feature configurations for each attack type: DIS-Flooding, Local Repair, and Worst Parent. Each value is computed over the complete dataset for the corresponding attack type, including all CSV files generated across all attack variants and network sizes.}
    \label{fig:f1_all}
\end{figure*}
\subsection{Effect of TX and RX Features on IDS Performance}
To systematically evaluate the contribution of TX and RX features to IDS detection performance, we design an ablation study comparing five feature configurations. The baseline configuration uses only RPL features, comprising 14 features constructed from the mean and standard deviation of seven routing-layer statistics: Rank, DIO sent, DIO received, DIS sent, DIS received, DAO received, and total messages. The remaining four configurations are: (i) \textit{All features} (RPL + TX + RX, 18 features), which evaluates the full augmented feature set; (ii) \textit{TX and RX only} (4 features), which isolates the discriminative power of radio features alone; (iii) \textit{All except RX} (RPL + TX, 16 features), which assesses the individual contribution of TX; and (iv) \textit{All except TX} (RPL + RX, 16 features), which assesses the individual contribution of RX. Together, these configurations allow us to determine whether TX and RX provide complementary information to RPL features, and whether one radio feature is more informative than the other.
Each configuration is evaluated across three attack types, three behavioural variants (base, on--off, and gradual change), and four network sizes ($N \in \{5, 10, 15, 20\}$), resulting in 12 experimental configurations per attack type. The mean F1-score is aggregated across all variants and network sizes for each attack type, given in \Cref{fig:f1_all}.

For \textbf{DIS-Flooding} (\Cref{fig:f1_second}), the TX and RX-only configuration achieves the highest score of 0.990, marginally outperforming the full feature set (0.986), the RPL-only baseline (0.984), and the two ablation configurations (0.986 and 0.985). The narrow performance gap across all configurations suggests that DIS-Flooding induces strong and consistent changes in both routing and radio activity, making it highly detectable regardless of the feature set used. For \textbf{Local Repair} (\Cref{fig:f1_third}), the TX and RX-only configuration achieves the highest F1-score of 0.999, substantially outperforming the full feature set (0.983), the RPL-only baseline (0.980), and the two ablation configurations (0.976 and 0.982). Notably, removing RX causes a larger performance drop (0.976) than removing TX (0.982), indicating that RX contributes more discriminative information than TX for this attack type. Local Repair attacks trigger repeated DODAG reconstruction cycles that generate bursts of control message transmissions, elevating reception activity at neighbouring nodes beyond normal levels --- a pattern that RX captures more effectively than RPL statistics alone. For \textbf{Worst Parent} (\Cref{fig:f1_fourth}), the TX and RX-only configuration again achieves the highest F1-score of 0.998, while the RPL-only baseline yields the lowest score of 0.957, indicating that routing-layer features alone are insufficient for detecting this attack. Removing RX causes a drop to 0.967, while removing TX yields a higher score of 0.975, confirming that RX is the more dominant radio feature for Worst Parent detection. The full 18-feature set scores 0.970, which is lower than the TX and RX-only configuration, suggesting that combining RPL and radio features introduces some redundancy that slightly reduces detection performance for this attack type.
Unlike RPL features, which capture routing protocol-level responses to attacks, TX and RX directly reflect node-level communication activity, providing a more immediate and sensitive signal. Attacks such as Worst Parent subtly manipulate routing decisions without strongly affecting RPL statistics, yet their impact on packet transmission and reception is directly measurable through TX and RX.

\section{Related Work} 
Machine learning-based IDS has been widely studied for RPL-based IoT networks, where sink-collected RPL control-traffic statistics serve as effective security signals. Kaveh et al. \cite{kaveh2024impact, kaveh2025factors} demonstrated that attack variations and topology changes significantly affect IDS generalizability, showing that models trained on one attack variation may perform poorly under another. Complementing this, Violettas et al. in \cite{violettas2021softwarized} proposed ASSET, a softwarized IDS for RPL networks combining anomaly-based and specification-based detection with centralized monitoring and attacker identification, highlighting the value of cross-layer visibility in constrained IoT environments. From a broader attack coverage perspective, Garcia Ribera et al. \cite{garciaribera2022intrusion} designed a hybrid IDS that targets multiple RPL attacks, including Blackhole, DIS, and version number attacks, and notably analysed detection overhead in terms of CPU usage and transmission/reception activity \cite{garciaribera2022intrusion}. Prajapati et al. in \cite{prajapati2025comprehensive,bang2022assessment} proposed a taxonomy of attack categories and discussed defence mechanisms, datasets, and open challenges. Despite these advances, existing studies rely predominantly on routing-layer features, providing only a partial view of network activity. In \cite{canbalaban2020cross}, the authors showed that incorporating link-layer features alongside routing statistics can reduce false positives and improve detection in RPL-based IoT networks. 

However, radio-level features such as TX and RX have not been systematically incorporated into learning-based IDS models. Our work addressed this gap and evaluated its impact on LSTM-based intrusion detection.

\section{Conclusions and Future Scope}
In this paper, we augmented the RPL feature set with Transmit (TX) and Receive (RX) radio features and evaluated their impact on LSTM-based IDS performance across three attack types, three behavioural variants, and network sizes of 5 to 20 nodes. The TX and RX-only configuration achieves the highest F1-score across all three attack types, and the \textit{All except TX} configuration consistently outperforms the \textit{All except RX} configuration, confirming that receiver-side communication activity is a more discriminative signal for RPL-based attack detection. These results demonstrate that TX and RX provide discriminative information beyond RPL routing-layer statistics, making them effective features for building IDS.

As future work, we plan to incorporate RSSI and LQI features, conduct feature importance analysis, and evaluate the cross-attack generalizability of the trained models.

\section*{Acknowledgment}
This research has been supported by the Swedish Governmental Agency for Innovation Systems (VINNOVA) through the project Robust IoT Security: Intrusion Detection Leveraging Contributions from Multiple Systems (2023-02982).

\bibliographystyle{IEEEtran}
\bibliography{bibtex/bib/bibliography} 
\end{document}